\documentclass[twocolumn,showpacs,amsmath,prl,superscriptaddress]{revtex4-1}
\usepackage{color}
\usepackage{graphicx}
\usepackage{epsfig,psfrag}
\usepackage{amsmath}
\usepackage{amssymb}
\usepackage{amsbsy}
\usepackage{amsthm}
\usepackage{amsfonts}
\usepackage{wasysym}
\usepackage{bbm}
\usepackage{tabularx}
\usepackage{euscript}
\usepackage{enumerate}
\usepackage{amsfonts}
\usepackage{exscale}
\usepackage{bbold}
\usepackage{float}
\usepackage{slashed}
\usepackage{subfigure}
\usepackage{comment}
\usepackage[colorlinks,citecolor=blue]{hyperref}

\newcommand{\bea}{\begin{eqnarray}}
\newcommand{\eea}{\end{eqnarray}}
\newcommand{\ba}{\begin{array}}
\newcommand{\ea}{\end{array}}
\newcommand{\B}{\textcolor{blue}}

\def\bea{\begin{eqnarray}}
\def\eea{\end{eqnarray}}
\def\Tr{\mathrm{Tr}}
\def\nn{\nonumber}
\def\bea{\begin{eqnarray}}
\def\eea{\end{eqnarray}}
\def\nn{\nonumber}
\def\ba{\begin{array}}
\def\ea{\end{array}}
\def\nn{\nonumber}
\def\Tr{\text{Tr}}
\def\sgn{\text{sgn}}

\begin{document}

\title{An exponential ramp in the quadratic Sachdev-Ye-Kitaev model}

\author{Michael Winer}
\affiliation{Condensed Matter Theory Center and Joint Quantum Institute, Department of Physics, University of Maryland, College Park, Maryland 20742, USA}

\author{Shao-Kai Jian}
\affiliation{Condensed Matter Theory Center and Joint Quantum Institute, Department of Physics, University of Maryland, College Park, Maryland 20742, USA}

\author{Brian Swingle}
\affiliation{Condensed Matter Theory Center and Joint Quantum Institute, Department of Physics, University of Maryland, College Park, Maryland 20742, USA}

\begin{abstract}
A long period of linear growth in the spectral form factor provides a universal diagnostic of quantum chaos at intermediate times. By contrast, the behavior of the spectral form factor in disordered integrable many-body models is not well understood. Here we study the two-body Sachdev-Ye-Kitaev model and show that the spectral form factor features an exponential ramp, in sharp contrast to the linear ramp in chaotic models. We find a novel mechanism for this exponential ramp in terms of a high-dimensional manifold of saddle points in the path integral formulation of the spectral form factor. This manifold arises because the theory enjoys a large symmetry group. With finite nonintegrable interaction strength, these delicate symmetries reduce to a relative time translation, causing the exponential ramp to give way to a linear ramp.
\end{abstract}

\maketitle 

\B{\it Introduction.}---Among the many characteristics of quantum chaotic systems~\cite{haake2010quantum}, random-matrix-like energy levels provide a particularly universal signature~\cite{PhysRevLett.52.1,mehta2004random,doi:10.1063/1.1703775,wigner1959group, bohigas1984chaotic, brezin1997spectral, guhr1998random, heusler2004universal}. Indeed, it is widely accepted that random-matrix-like energy levels can be taken to define quantum chaos. This definition is a good one because it gives a simple universal characterization and because many systems that are intuitively chaotic in fact exhibit random-matrix-like energy levels.

Such a characterization of quantum chaos is well-established for single-body systems~\cite{berry1977level, PhysRevLett.42.1189,Casati1980,Berry1981,doi:10.1098/rspa.1985.0078} and, more recently, for a variety of many-body systems~\cite{dubertrand2016spectral, Cotler2017, cotler2017chaos, dyer20172d, you2017sachdev, garcia2017analytical, gharibyan2018onset, Stanford2018, Bertini_2018, liu2018spectral, okuyama2019spectral}. However, the crossover between single-body chaos and many-body chaos is less well understood. In this paper, we address this problem by characterizing the energy level statistics of a many-body model composed of non-interacting fermions filling random matrix single-particle energy levels. This is a necessary first step to studying how full many-body chaos is established in a weakly interacting gas with chaotic single-particle dynamics.

\begin{figure}
    \centering
    \includegraphics[width=5cm]{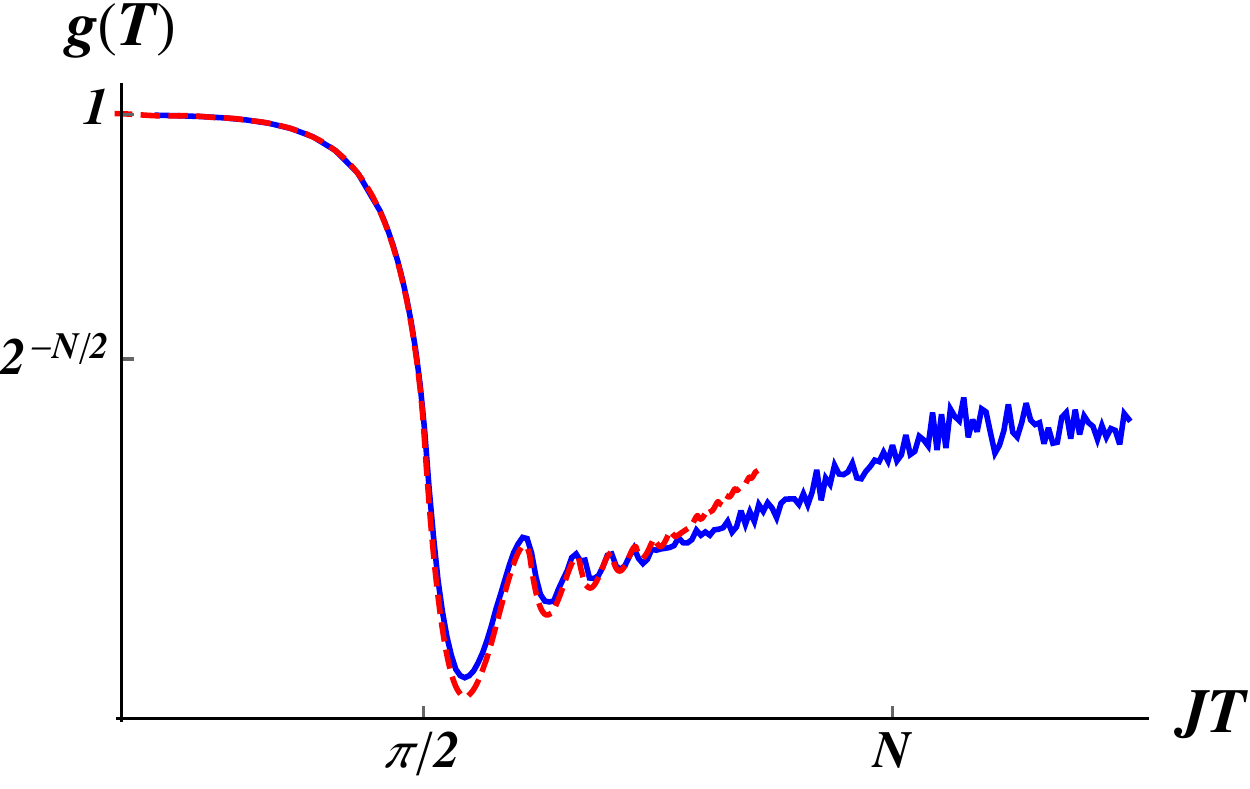}
    \caption{The normalized spectral form factor of the two-body SYK (SYK$_2$) model . The blue solid (red dashed) line represents the numerical (theoretical) curve of the spectral form factor. The theoretical curve is given by Eq.~(\ref{eq:SFF}), and the numerical data is drawn from $N=200$ SYK$_2$ model averaged over $10^4$ samples. The deviation at large $JT$ is due to the mass of soft modes being suppressed at $JT \sim N$.}
    \label{fig:sff}
\end{figure}

Focusing on infinite temperature in a system of $N$ Majorana fermions, we formulate our analysis in terms of the many-body spectral form factor (SFF)~\cite{brezin1997spectral, Cotler2017}. If the microscopic timescale is $J^{-1}$, then at early time of order $J^{-1}$, the spectral form factor is controlled by the density of states of the system and is not sensitive to single-particle level correlations. At times long compared to $J^{-1}$ but not scaling with $N$, we find that the spectral form grows exponentially with time, a feature we call an exponential ramp. This should be contrasted with the linear ramp observed for conventional many-body chaos. At times of order $N$ or longer, the spectral form factor shows a plateau~\cite{liao2020manybody}, but we do not study this regime in our work. A prototype of the SFF of disordered integrable system is shown in Fig.~\ref{fig:sff}. We note that~\cite{liao2020manybody} appeared shortly before our work and studies the same problem, though mostly using different techniques.

The infinite temperature SFF is given by the disorder average of the magnitude squared of the trace of the time evolution operator,
\begin{equation}
    \text{SFF}(T) = \langle |\text{Tr}[U(T)]|^2\rangle_{\text{disorder}}.
\end{equation}
We compute this quantity for a system of real fermions with single-particle energy levels given by the eigenvalues of a random antisymmetric Hermitian matrix. Note that the single-particle eigenvalues come in plus-minus pairs and the system can be taken to be time-reversal invariant. We choose to call this many-body system the two-body Sachdev-Ye-Kitaev model (SYK$_2$), but it has appeared under a variety of names in the literature~\cite{maldacena2016remarks, banerjee2017solvable, Cotler2017, jian2017solvable, nosaka2018thouless, garcia2018chaotic, garcia2018exact, garcia2019many, flores2001spectral, dai2019global, aref2019replica, lau2019randomness, lau2020correlated, garcia2018chaotic}. 

Our analysis is based on a path integral formulation of the SFF involving two replicas of the system with identical randomness. In the large $N$ limit and for times short compared to $N$, the resulting path integral may be analyzed via saddle point and steepest descent methods. The action governing the SFF is formulated in terms of fields depending on two times, and the resulting theory possesses a very large symmetry group. We find that the theory exhibits a time-dependent pattern of symmetry breaking leading to a vacuum manifold of replica-nondiagonal solutions which yield an exponential ramp,
\begin{equation}
    \log \text{SFF} = \frac{JT}{\pi} \log N + \cdots.
\end{equation}

The remainder of this paper is organized as follows. We first setup the general path integral formalism for the SFF with arbitrary $q$-body SYK interactions. Then we analyze the $q=2$ case in detail. By using saddle point analysis, we obtain both the slope and the novel exponential ramp. The exponentially growing vacuum manifold explains the corresponding exponential ramp. Finally, we comment on the effects of weak interactions that ultimately restore the linear ramp. We also provide an explicit evaluation of the SFF of the SYK$_2$ model restricted to time-translation invariant fluctuations in the Supplemental Material~\cite{supp}, and obtain consistent results.

\B{\it Model and path integral.}---We consider the SYK model and represent the SFF via a path integral. The Hamiltonian of the SYK$_q$ model with general $q$-body interaction is
\begin{equation}
    H[\psi] = i^{q/2}\sum_{1 \leq j_1 <...< j_q \leq N} J_{j_1j_2...j_q}\psi^{j_1}\psi^{j_2}...\psi^{j_q},
\end{equation}
where $\psi^i, i=1,...,N$ represents the Majorana fermions and satisfy the anticommutation relation $\{\psi^i,\psi^j\}=\delta_{ij}$, and $J_{j_1...j_q}$ is a Gaussian variable (the disorder) with mean zero and variance $\langle J_{j_1...j_q}^2 \rangle=\frac{J^2 (q-1)!}{N^{q-1}}$.
We analyze this model using large $N$ to control the calculation.

The SFF can be thought of as the thermal partition function with imaginary temperature which translates into conventional real time. We are interested in the following normalized SFF,
\bea
    g(T) \equiv \frac{\langle |Z(i T)|^2 \rangle_{\rm disorder}}{\langle |Z(0)|^2 \rangle_{\rm disorder}},
\eea
where $Z$ and $T$ represent the partition function and the time, respectively. The norm $|Z(i T)|^2=Z(i T) Z(i T)^\ast$ is to make the SFF real. The partition function is represented as a Grassman path integral,
\bea
    Z(i T) = \int D \psi \exp\int dt \Big[- \frac12 \psi \partial_t \psi + i  H[\psi]  \Big].
\eea
Thus the path integral representation of the SFF~\cite{Stanford2018} is
\bea
    g(T) &=& \int D\psi_L D\psi_R \exp \int \Big[- \frac12 \psi_\alpha \partial_t \psi_\alpha+ i (H_L - H_R) \Big], \nn \\
     H_L &=& H[\psi_L], \quad \quad H_R=H[\psi_R].
\eea

It is straightforward to integrate out the Gaussian random variable $J_{j_1 j_2 ... j_q}$ and get the effective action. In terms of the bilocal field $G_{\alpha\beta}(\tau_1,\tau_2)= \frac 1N\sum_i \psi^i_\alpha(\tau_1)\psi^i_\beta(\tau_2)$, and the self-energy $\Sigma_{\alpha\beta}(\tau_1,\tau_2)$, the effective action is given by
\bea\label{eq:Q-action}
    g(T) &=& \int DG D\Sigma \exp N \Big[ \frac12 \Tr \log( \partial_t - i \hat\Sigma) \nn\\
    && + \frac{1}2 \int dt_1 dt_2 (i \Sigma_{\alpha \beta} G_{\alpha\beta} - \frac{J^2 }{q}  (-1)^{\alpha+\beta} G_{\alpha\beta}^q ) \Big] , 
\label{eq:Action}
\eea
where a hat above a variable signals a matrix representation, $(\hat \Sigma)_{\alpha\beta} \equiv \Sigma_{\alpha\beta}$. Due to the antiperiodic boundary conditions on the fermions, both $G_{\alpha \beta}$ and $\Sigma_{\alpha \beta}$ are antiperiodic under time shifts by $T$. 

In the case of the SYK$_2$ model, i.e., $q=2$, we can further integrate out $G$ to get
\bea\label{eq:2-action}
    g(T) = \int D \Sigma \exp N \Big[ && \frac12 \Tr \log (\partial_t - i \hat\Sigma) \nn \\
    && - \frac1{4J^2} \int (-)^{\alpha+\beta} \Sigma_{\alpha\beta}^2(t_1, t_2) \Big].
\eea
Owing to the nice large-$N$ structure in Eq.~(\ref{eq:2-action}), we use saddle point approximation to evaluate the path integral in the following.

\B{\it The slope.}---The equation of motion from Eq.~(\ref{eq:2-action}) is
\bea\label{eq:EOMq=2}
     (i \partial_t +\bar \Sigma)_{\alpha\gamma} \ast (-)^{\gamma+\beta} \bar \Sigma_{\gamma\beta} = J^2 \delta_{\alpha\beta}
\eea
where $A \ast B \equiv \int dt A(t_1, t) B(t, t_2)$ is a matrix product in time arguments, which are also suppressed. Assuming time-translation symmetry and going to frequency space, the replica-diagonal solution is given by
\bea\label{eq:solution}
    \bar\Sigma(\omega) = \begin{cases} \frac12(\omega-\sqrt{\omega^2 - 4J^2}), & \quad   \omega > 2J \\
                                    \frac12(\omega - i \sqrt{4J^2 - \omega^2} \sigma_z ), & \quad 0< \omega < 2J
                    \end{cases}
\eea
where $\sigma_\mu$ denote Pauli matrices, and the solution for $\omega<0$ can be obtained through the relation $\bar\Sigma(-\omega) = - \bar\Sigma(\omega)^T $. Actually, there are four solutions for Eq.~(\ref{eq:EOMq=2}) at each frequency, however one can show that above choice Eq.~(\ref{eq:solution}) is the dominant one~\cite{supp}.

\begin{figure}
\includegraphics[width=5cm]{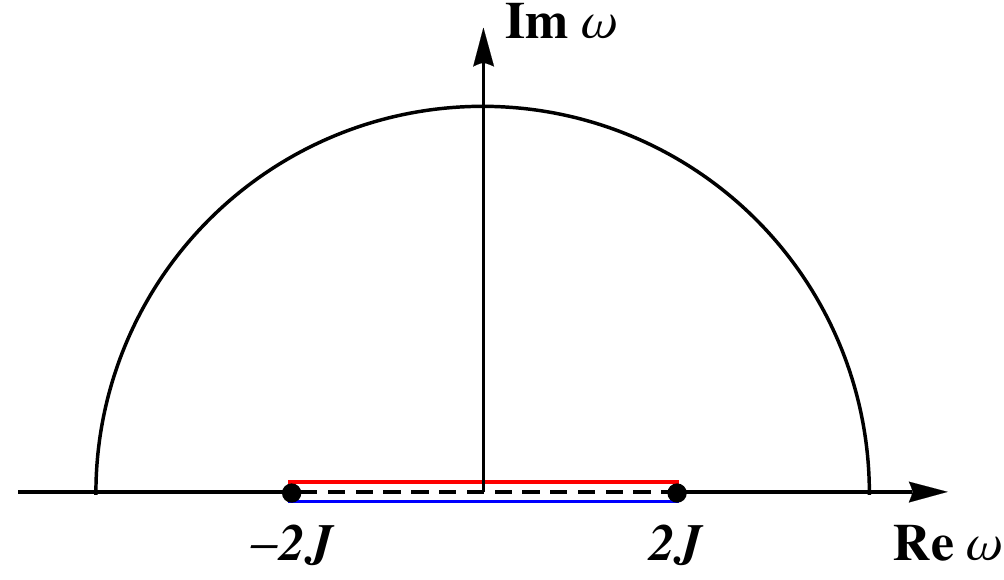}
\caption{The contours used to evaluate Eq.~(\ref{eq:contour}). The integral picks up the difference between the two contours going above and below the branch cut, which is twice the imaginary part of the saddle point solution.\label{fig:contour}}
\end{figure}

Since $T$ and $J$ are the only dimensionful quantities in the problem, the $T$ dependence in the SFF must be through $J T$. Hence, to access time derivatives, one can also differentiate Eq.~(\ref{eq:2-action}) with respect to $J$, 
\begin{equation}
\frac{d S_{\rm saddle}}{d(J^{-2})}= -\frac{N}4 \int dt_1 dt_2 \sum_{\alpha,\beta}(-)^{\alpha+\beta}\bar \Sigma_{\alpha\beta}(t_1,t_2)^2.
\end{equation}
The right hand side can be evaluated to leading order in $N$ by plugging in the saddle-point solution. The equation of motion, Eq.~(\ref{eq:EOMq=2}), then implies
\bea
 \int dt_1 dt_2 \sum_{\alpha,\beta}(-)^{\alpha+\beta} \bar\Sigma_{\alpha\beta}(t_1,t_2)^2 &=& i T \sum_\alpha \partial_{t} \bar \Sigma_{\alpha\alpha}(t)|_{t\to 0^+}, \nn
\eea
Consider first the limit $T=\infty$, 
\bea\label{eq:contour}
\sum_\alpha \bar \Sigma_{\alpha \alpha}(t)|_{T=\infty} = \int_{-\infty}^\infty \frac{d \omega}{2\pi}  e^{i\omega t} \sum_\alpha \bar \Sigma_{\alpha\alpha}(\omega).
\eea
The frequency space integrand has a branch cut from $-2J$ to $+2J$, as indicated in Eq.~(\ref{eq:solution}). The integral is defined by taking $\bar \Sigma_{11}(\omega)$ to be above the branch cut and $\bar \Sigma_{22}(\omega)$ below the branch cut. For $t>0$, the integrand goes to zero in the upper half of the complex plane, and we can extend the integral to cover the contour shown in Fig.~\ref{fig:contour}. As a result, Equation~(\ref{eq:contour}) is equal to the integral of the discontinuity over the branch cut, i.e.,
\bea
    i\sum_{\alpha} \bar \Sigma_{\alpha \alpha}(t)|_{T=\infty} &=&  \int_{-2J}^{2J} \frac{d \omega}{2\pi}  e^{i\omega t}\sqrt{4J^2-\omega^2} \\  
    &=& J^2 \frac{2 J_1(2 Jt)}{2Jt}\sgn (t),
\eea
where $J_1$ is a Bessel function of the first kind. 

For finite $T$, the antiperiodic boundary condition can be accounted for by summing over translates, 
\bea
    && \int dt_1 dt_2 \sum_{\alpha\beta }(-1)^{\alpha+\beta}\Sigma_{\alpha\beta}(t_1,t_2)^2 \\
    &=& TJ^2 \partial_t \sum_k (-1)^k \frac{2 J_1(2 J(t+kT))}{2J(t+kT)}\sgn (t+kT)|_{t=0^+}, \nn
\label{eq:AltSum}
\eea
This formula can be manipulated to get 
\begin{equation}
\frac{d \log g(T)}{d T}=\partial_T \frac{-N}{2}\sum_{k\neq 0} \frac{(-1)^k}{k} \frac{2 J_1(2 JkT)}{2JkT}\sgn(k).
\label{eq:logDiff}
\end{equation}
By taking into account of the initial condition of normalized SFF, i.e., $g(0) = 1$, this can be massaged into the final form
\begin{equation}
    \log g(T)=- N \log 2- N \sum_{k=1}^\infty \frac{(-1)^k}{k} \frac{2 J_1(2 JkT)}{2JkT}.
    \label{eq:slopeFinal}
\end{equation}
For $JT \ll 1$, the alternating sum in Eq.~(\ref{eq:slopeFinal}) 
evaluates to~\cite{supp} 
$\log g(T)= -N\frac{J^2T^2}{8}$.
While for large $JT \gg 1$ we expect the sum goes to zero as $T^{-3/2}$.

\B{\it Spontaneous symmetry breaking.}---The leading order in $N$ saddle point analysis explains the slope of the SFF. The exponential ramp of the SFF is due to fluctuations around saddle point. In particular, due a time-dependent pattern of symmetry breaking, there are a growing number of fluctuations which are not suppressed at large $N$.

The equation of motion Eq.~(\ref{eq:EOMq=2}) has a very peculiar symmetry~\cite{supp}. Namely, at the fixed frequencies $\pm \omega$, it is invariant under an SU(2) conjugation,
\bea
\hat \Sigma(\pm \omega) \rightarrow U^\dag_{\pm \omega} \hat \Sigma(\pm \omega) U_{\pm \omega},
\eea 
where $U_{\pm \omega}$ denotes an SU(2) matrix and $U_{\pm \omega}$ are related by $U_{ \omega}=\sigma_z U_{-\omega}^* \sigma_z$. 
Note that the SU(2) transformation at $-\omega$ is locked to the transformation at $\omega$. From the saddle point solution Eq.~(\ref{eq:solution}), it is clear that the SU(2) symmetry is not broken for $|\omega|>2J$; while for $0<|\omega|<2J$, the SU(2) symmetry is spontaneously broken to $O(2)$.Thus, for each pair of frequencies $|\omega|<2J$, the vacuum manifold is $SU(2)/O(2) \sim S^3/S^1 \sim S^2$. Due to the antiperiodic boundary conditions, the allowed Matsubara frequencies are $\omega_n = \frac{n \pi}{T}$ with odd integer $n$, and the vacuum manifold is the tensor product of each pair of Matsubara frequency in $|\omega_n|<2J$, i.e., $(S^2)^{\otimes [\frac{J T}\pi]}$, $[\frac{J T}\pi]$ denotes the nearest integer to $\frac{J T}\pi$.

\begin{figure}
    \centering
    \includegraphics[width=5cm]{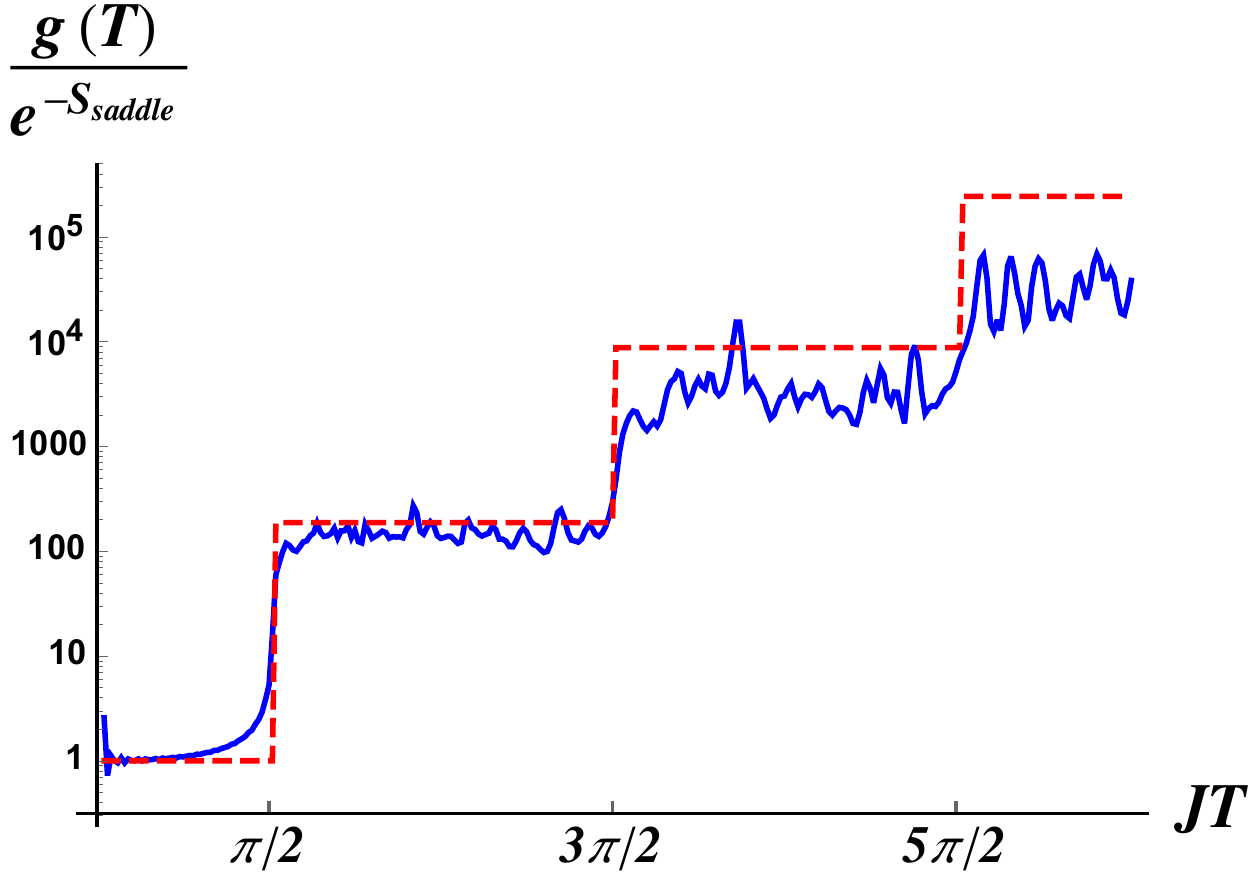}
    \caption{The discrete jumps of the spectral form factor from the theory Eq.~(\ref{eq:SFF}) and the numerics. The spectral form factor is divided by the continuous saddle point value to make the jumps visible.}
    \label{fig:jump}
\end{figure}

A convincing evidence of the vacuum manifold $(S^2)^{\otimes [\frac{J T}\pi]}$ is from the fact that the size of this manifold is fundamentally discontinuous in $JT$, jumping every time $JT$ crosses a half-multiple of $\pi$. This can be seen from the discrete jumps of the SFF plotted in Fig.~(\ref{fig:jump}), where the $S_{\rm saddle}$ has been subtracted to make the jumps visible.

\B{\it The ramp.}---The vacuum manifold $(S^2)^{\otimes [\frac{J T}\pi]}$ increases exponentially as time increases, leading to an exponential ramp. Especially, the zero mode fluctuation in each dimension of the vacuum manifold $S^2$ is $\sqrt N$ times bigger than a massive mode. This implies the leading order behavior of the ramp is given by
\begin{equation}
g(T) \propto N^{\frac{JT}\pi}.
\end{equation}

To make this more rigorous, we now calculate the saddle point fluctuation at $J T \gg 1$. By varying around the saddle point, $\bar \Sigma(\bar \omega) \rightarrow \bar \Sigma(\bar \omega) + \delta \Sigma(\omega_1, \omega_2)$, $\bar \omega = \frac{\omega_1 + \omega_2}2$, the effective action at quadratic order is
\bea
    \frac{\delta S}{N} &=& \frac1{4J^2T^2} \sum_{\alpha,\beta, \omega_{1,2}} K_{\alpha\beta}(\omega_1, \omega_2) |\delta \Sigma_{\alpha\beta}(\omega_1, \omega_2)|^2,
\eea
where the sum is running over two contours $\alpha, \beta = 1, 2$ and two Matsubara frequencies $\omega_{1,2}$, and the kernel is given by 
\bea
K_{\alpha\beta}(\omega_1, \omega_2)=(-)^{\alpha+\beta}- \frac1{J^2} \bar \Sigma_{\alpha\alpha}(\omega_1) \bar \Sigma_{\beta\beta}(-\omega_2).
\eea

When $\alpha \neq \beta$ and $-2J<\omega_1=\omega_2<2J$, the expression becomes zero because of the vacuum modes. At the naive quadratic order, this means the volume of the saddle point diverges. But to higher orders in the action, we know that the vacuum manifold is not an infinite plane, but rather a $S^2$ with 
surface area 
$4\pi (J^2- \frac{\omega^2}4)$. This is as opposed to an area of $ \frac{2\pi}N J^2$ if we had ignored the $\Tr \log$ term in Eq.~(\ref{eq:2-action}). 
So including the saddle point fluctuations, the SFF is given by
\bea
    g(T) &=& e^{-S_{\rm saddle}} \int D\Sigma e^{-\delta S}\\
     &=& e^{-S_{\rm saddle}} \prod_{<\alpha,\beta,\omega_{1,2}>}  K_{\alpha\beta}(\omega_1, \omega_2)^{-1/2} \\
     && \times\prod_{\omega_1=\omega_2=\omega}^{0<\omega <2J} N\frac{4\pi (J^2- \frac{\omega^2}4)}{2\pi J^2}.
\eea
where in the second line we integrate over the massive Gaussian fluctuations and properly normalize it by dividing a free theory, e.g., a theory without the $\Tr \log$ term in Eq.~(\ref{eq:2-action}). The angle brackets in the product mean exclusion of zero modes. The last factor is to account for the zero modes, which span over the vacuum manifold $(S^2)^{\otimes [\frac{J T}\pi]}$. Since the saddle point action has already been evaluated in the previous sections, our goal is to evaluate the following quantity, 
\bea\label{eq:productF}
    \mathcal{F} &=& \log \prod_{<\alpha,\beta,\omega_{1,2}>}  K_{\alpha\beta}(\omega_1, \omega_2)^{-1/2} \\
    &=& - \frac12 \sum_{<\alpha,\beta,\omega_{1,2}>} \log K_{\alpha\beta}(\omega_1, \omega_2).
\eea
While it is unclear how to exactly evaluate $\mathcal F$, one can prove that it converges~\cite{supp}, and solve it in the $JT\gg 1$ limit.

\B{\it Evaluation of Eq.~(\ref{eq:productF}).}---At $JT\gg 1$ limit, the discreteness of Matsubara frequency is small, so the following decomposition is a good approximation, $\mathcal{F} = \delta \mathcal{F} - \frac12 \sum_{\alpha,\beta} \int_{\omega_1,\omega_2} \log K_{\alpha\beta}(\omega_1, \omega_2)$, in which $\delta \mathcal F$ is the difference between the sum and the integral,
\bea \label{eq:dF}
    \delta \mathcal F \equiv \mathcal{F} + \frac12 \sum_{\alpha,\beta} \int_{\omega_1,\omega_2} \log K_{\alpha\beta}(\omega_1, \omega_2).
\eea
Fortunately, it can be shown that the integral $\sum_{\alpha,\beta} \int_{\omega_1,\omega_2} \log K_{\alpha\beta}(\omega_1, \omega_2)$ actually vanishes~\cite{supp}. 

Therefore all the contributions come from $\delta \mathcal F$, i.e., from the difference between the sum and the integral. In particular, the soft modes near zero modes dominate the contribution. Namely, for $\omega_{1,2} = \omega \pm \frac{\Omega_k}{2}$ and $|\omega|< 2J$, soft modes for small $\Omega$ dominate $\delta \mathcal F$. The spacing between $\bar \Sigma(\omega)$ of adjacent Matsubara modes is $\delta \bar\Sigma(\omega) = 2 \partial_\omega \bar \Sigma(\omega) =  \frac{\pi}T(1 \pm i\frac{\omega}{\sqrt{4J^2-\omega^2}})$, which has a magnitude of 
\bea 
    \Delta(\omega) \equiv |\delta \bar\Sigma(\omega)| =\frac{\pi}{T}\frac{2J}{\sqrt{4J^2-\omega^2}}.
\eea 
We cut off the sum at any integer $M$ such that $1\ll M \ll JT$ and cut off the integral at $M+1/2$, and approximate the inverse kernel as linear functions in $\Omega$ within this region. The difference is thus
\bea
&& \delta F = -2\times2 \times \frac {1}{2} \times \\
&& \sum_{0<\omega<2J} \left( \sum_{k=1}^M \log \frac{k\Delta(\omega)}{J}-\int_0^{M+\frac 12}\log \frac {u\Delta(\omega)}J du\right),
\eea
where $\Omega_k = \frac{2k \pi}{T}$ and $k$ is an integer. In the first line, we include two factors of two from the two contours $\alpha \ne \beta$ and from both the positive and negative values of $k$ in the sum, respectively. 

For $JT \gg 1$, using the Stirling's approximation for the sum, we have 
\begin{equation}\label{eq:dF2}
\delta \mathcal F \approx \sum_{0< \omega < 2J} \log \frac{\Delta(\omega)}{2\pi J}
= \frac{J T}\pi \log \frac{e}{4 J T}.
\end{equation}
The mass of soft modes is proportional to $\frac1{JT}$, and leads to the $JT$ factor insides the logarithm. 

Combining with the essential factor from the vacuum manifold, i.e.,
\bea
    \sum_{0<\omega<2J} \log N\frac{4\pi (J^2- \frac{\omega^2}4)}{2\pi J^2} \approx \frac{JT}\pi \log \frac{32 N}{e^2},
\eea
where the discrete sum is approximated by an integral, we have the SFF at $JT \gg 1$,
\bea\label{eq:SFF}
    \log g(T) + S_{\rm saddle}=\frac{JT}{\pi} \left(\log \frac{N}{JT} + \log\frac{8}e \right).
\eea
The $JT$ factor insides the logarithm indicates that the exponential ramp terminates at $J T \sim N$. This correction is due to the soft mode as we have seen from $\delta F$ in Eq.~(\ref{eq:dF2}). At time scales $JT\gg N$, the soft mode is as important as the zero mode, and these fluctuations together gives the plateau in the SFF as shown in Fig.~\ref{fig:sff}.

\textcolor{blue}{\it Discussion.}---Besides providing a novel explanation of the exponential ramp, the symmetry structures can also help to understand the transition to a linear ramp.
As a many-body chaotic Hamiltonian features a linear ramp, it is natural to expect that a finite interaction strength will explicitly break the SU(2) symmetry at each frequency down to a residual relative time translation~\cite{Stanford2018} that is defined by $U_a \equiv e^{i \omega a \sigma^z}$, where $a$ is a frequency-independent parameter.
To see this, one can consider, for example, the equation of motion from Eq.~(\ref{eq:Q-action}),
\bea
    \hat G(\omega) &=& [i \omega \hat 1 - i \hat \Sigma(\omega)]^{-1} , \\
    i \Sigma_{\alpha\beta}(\omega) &=& (-1)^{\alpha+ \beta} \Big[ J_2^2 G_{\alpha\beta}(\omega) \nn\\
    && + J_q^2 2\pi \delta \Big(\omega- \sum_{j=1}^{q-1} \omega_j \Big) \prod_{j=1}^{q-1}G_{\alpha\beta}(\omega_j)  \Big],
\eea
where we have kept the $q=2$ part as well, i.e., $J_2$ and $J_q$ represent the effective interaction strengths of SYK$_2$ and SYK$_q$ models, respectively.
It is apparent that at any finite $J_q$ the above equation of motion is only invariant under the relative time translation $U_a$.
Consequently, the vacuum manifold is reduced from $(S^2)^{\otimes [JT/\pi]}$ down to $a \in S^1$ with the identification $a = a + 2T$. And it increases linearly in time (i.e., the circumference increases linearly in time), in sharp contrast to the exponential increase for the $q=2$ case, causing a transition from an exponential ramp to a linear ramp~\cite{supp}.

\B{\it Conclusions.}---In this paper, we studied the spectral form factor (SFF) of the quadratic SYK model, and found an interesting exponential ramp in sharp contrast to the linear ramp in chaotic models. This exponential ramp can be understood as arising from a high-dimensional manifold of saddle points, which result from a huge number of delicate symmetries present in the quadratic SYK model. Because the SFF in the quadratic SYK model is ultimately equivalent to that of a Gaussian random matrix theory, our result not only implies that all noninteracting disordered systems feature an exponential ramp, but also gives a universal explanation for this phenomena. Moreover, our mechanism also indicates that a dramatic change of manifolds underlines the single-to-multiparticle chaos transition in the SFF when nonintegrable interactions are turned on. It is interesting to explore this transition in the path integral language in more detail, which we leave as future work.

\B{\it Acknowledgements.}---We thank Victor Galitski, Yunxiang Liao, and Amit Vikram for helpful discussions. This work is supported in part by the Simons Foundation via the It From Qubit Collaboration (S. K. J. and B. S.) and by the Air Force Office of Scientific Research under award number FA9550-17-1-0180 (M.W.). M.W. is also supported by the Joint Quantum Institute

\bibliography{refs.bib}







.

\newpage
\begin{widetext}

\section{Supplemental Material}
\renewcommand{\theequation}{S\arabic{equation}}
\setcounter{equation}{0}
\renewcommand{\thefigure}{S\arabic{figure}}
\setcounter{figure}{0}
\renewcommand{\thetable}{S\arabic{table}}
\setcounter{table}{0}

\subsection{A. More Details on Equation~(\ref{eq:logDiff})}
The slope portion of the SFF is given by Eq.~(\ref{eq:logDiff}), reproduced here in integrated form:
\begin{equation}
    \log g(T)=-N \sum_{k=1}^\infty \frac{(-1)^k}{k} \frac{2 J_1(2 JkT)}{2JkT}\sgn(k) - N\log 2.
\end{equation}
where we fix the integration constant by requiring $g(0) = 1$.
This infinite sum is absolutely convergent since it goes with $k$ as $k^{-5/2}$. It can also be approximated well in the large or small $JT$ limits.
In general for any function $f$, we can write, $\sum_{k=1}^\infty \frac {(-1)^k}{k} f(k)=\sum_{i=0}^\infty\sum_{k=1}^\infty \frac {(-1)^k}{k} f^{(i)}(0)\frac{k^i}{i!}=\lim_{x\to 0}\sum_{i=0}^\infty \partial_x^i \log \frac{1}{1+e^{-x}} \frac{f^{(i)}(0)}{i!}$. 
For a slow-varying function like $f(k)=\frac{2 J_1(2 JkT)}{2JkT}$, $JT\ll 1$, this last series converges quickly. It gives
\begin{equation}
    \log g(T)=  -N\frac{J^2T^2}{8}.
\end{equation}

\subsection{B. The symmetry of the action}

For reference, here's the action again:
\begin{equation}
S=N\left(\frac 12 \log \det (\partial_t\delta_{\alpha \beta} - i\Sigma_{\alpha\beta})-\frac {1}{4J^2}\int d\tau_1 d\tau_2 (-1)^{\alpha+\beta}\Sigma_{\alpha\beta}(\tau_1,\tau_2)^q\right).
\end{equation}
Let's also write down the symmetry transformations again:
\begin{center}
\begin{tabular}{c|c}
$\Sigma_{\alpha \beta}(\omega)$&$\Sigma_{\alpha \beta}(-\omega)$\\
\hline
$\begin{pmatrix}e^{i\theta}&0\\0&e^{-i\theta}\end{pmatrix}$&$\begin{pmatrix}e^{-i\theta}&0\\0&e^{i\theta}\end{pmatrix}$\\
$\begin{pmatrix}\cos\theta&i\sin\theta\\i\sin\theta&\cos\theta\end{pmatrix}$&$\begin{pmatrix}\cos\theta&i\sin\theta\\i\sin\theta&\cos\theta\end{pmatrix}$\\
$\begin{pmatrix}\cos\theta&\sin\theta\\-\sin\theta&\cos\theta\end{pmatrix}$&$\begin{pmatrix}\cos\theta&-\sin\theta\\\sin\theta&\cos\theta\end{pmatrix}$
\end{tabular}
\end{center}
The $\Tr \log$ term has to remain invariant under this transformation, because traces are always invariant under conjugation. We just need to worry about the $\Sigma^2$ term. 

The $\Sigma^2$ term is proportional to 
\begin{equation}
S_{\Sigma^2}=\Tr \sigma_z \Sigma(\omega)\sigma_z \Sigma^T(-\omega)
\end{equation}
Where all traces and matrix multiplication are over $\alpha \beta$ indices. The $\sigma_z$s are to enforce the signs. The transpose (not Hermitian conjugate, but transpose) is because we are squaring each element of the matrix rather than multiplying $\Sigma_{12}$ by $\Sigma_{21}$. If we conjugate  $\Sigma(\omega)$ and $\Sigma(-\omega)$ by $U_+$ and $U_-$ respectively we get 
\begin{equation}
S'_{\Sigma^2}=\Tr \sigma_z U_+\Sigma(\omega)U_+^\dagger\sigma_z U_-^*\Sigma^T(-\omega)U_-^T
\end{equation}
Using cyclicity of the trace we can turn this into
\begin{equation}
S'_{\Sigma^2}=\Tr \Sigma(\omega)U_+^\dagger\sigma_z U_-^*\Sigma^T(-\omega)U_-^T \sigma_z U_+
\end{equation}
If $U_+=\sigma_z U_-^* \sigma_z$ this works out to just
\begin{equation}
S'_{\Sigma^2}=\Tr \sigma_z \Sigma(\omega) \sigma_z \Sigma^T(-\omega)
\end{equation}
as desired.

\subsection{C. Goldstone modes}
There are two zero modes given by
\begin{equation}
\begin{split}
\delta_1 \Sigma(\omega)=\begin{pmatrix}
0 & i\\ -i &0
\end{pmatrix} \qquad
\delta_1 \Sigma(-\omega)= \begin{pmatrix}
0 & -i\\ i &0
\end{pmatrix}\\
\delta_2 \Sigma(\omega)=\begin{pmatrix}
0 & 1\\ 1 &0
\end{pmatrix} \qquad
\delta_2 \Sigma(-\omega)= \begin{pmatrix}
0 & 1\\ 1 &0
\end{pmatrix}
\end{split}
\end{equation}
In a purely quadratic expansion around a saddle point, a zero mode multiplies the action by a factor of infinity. This is because in a purely quadratic theory, a zero mode implies an infinite line of saddle points. But for this theory, we know non-perturbatively that the saddle points are given by
\begin{equation}
\begin{split}
\Sigma(\omega)=\begin{pmatrix}
\frac \omega 2+i\cos \phi\sqrt{J^2-\frac{\omega^2}{4}}&e^{i\theta}\sin \phi \sqrt{J^2-\frac{\omega^2}{4}}\\
e^{-i\theta}\sin \phi \sqrt{J^2-\frac{\omega^2}{4}} &\frac \omega 2-i\cos \phi\sqrt{J^2-\frac{\omega^2}{4}}
\end{pmatrix}\\
\Sigma(-\omega)=\begin{pmatrix}
-\frac \omega 2+i\cos \phi\sqrt{J^2-\frac{\omega^2}{4}}&e^{-i\theta}\sin \phi \sqrt{J^2-\frac{\omega^2}{4}}\\
e^{i\theta}\sin \phi \sqrt{J^2-\frac{\omega^2}{4}} &-\frac \omega 2+i\cos \phi\sqrt{J^2-\frac{\omega^2}{4}}
\end{pmatrix}
\end{split}
\end{equation}
Knowing the non-perturbative solution means that we can cut off the divergent integral due to the compact space of saddle points. We know that the sphere has surface area proportional to $J^2$, $N$ times more than what we would expect from a massive mode. That's why each of these spheres gives a factor of $N$.

\subsection{D. Convergence of the quantity $\mathcal F$}

\subsubsection{IR convergence}
When we have $|\omega|<2JT$, $|\Sigma(\omega)|=J$. This means that for $\omega_1-\omega_2$ small, and $|\omega_1|,|\omega_2|<2J$, we can have a very large contribution to the product when $\alpha\neq \beta$ and $\Sigma_{\alpha\alpha}=\overline{\Sigma_{\beta\beta}}$. But for any finite $JT$, there are only $O(J^2T^2)$ terms with $\omega_1,\omega_2$ within any $\delta \omega<<J$ of each other satisfying $|\omega|<2J$. This is a finite product, so can't give a divergent answer for any given $JT$. 

\subsubsection{UV convergence}
The issue of UV convergence is a little more complicated. For large $\omega$ we have $\Sigma(\omega)_{\alpha\alpha} \approx (-1)^{1+\alpha}J^2/\omega$. This means that the log of our product is in the UV limit
\begin{equation}
\log \textrm{Volume ratio}\approx \sum_{\alpha,\beta,\omega_1,\omega_2} (-1)^{\alpha+\beta} \frac{J^2}{\omega_1 \omega_2}    
\end{equation}
This sum is not absolutely convergent. However, there is a cancellation if we impose any finite UV cutoff (for instance a Lattice cutoff). This is the cancellation between $+\omega$ and $-\omega$, or alternatively between one contour and the other. Either of these cancellations is enough to cure UV divergences.

\subsection{E. Vanishing of the integral}

Is there an infinitely differentiable $f$ from $R$ to $C$ such that $\int_{-\infty}^\infty f^k(x) dx=0$ for all positive integers $k$? One might suspect the only answer is zero, which is true for discrete sums. But for integrals, one can plug in $f(x)=1/(x+i)^2$. We can run our contour around the upper half of the complex plane, and then use Cauchy's integral theorem to show that this will indeed give zero for any $k$.

From there, we can go further. For any polynomial $P$ such that $P(0)=0$ we have $\int_{-\infty}^\infty P(f(x))dx=0$. And finally, for any function $F$ such that $F(f(x))$ is analytic on either the top or bottom half of the complex plane and goes to zero sufficiently fast $\int_{-\infty}^\infty F(f(x))dx=0$. And, of course, this logic works just as well for double integrals and functions of two variables.

We have a whopping five problems when we try to apply this logic to our infinite sum of logs. 
\begin{enumerate}
    \item $\Sigma_{\alpha\alpha}$ isn't analytic, it has a branch cut between $\omega=-2J$ and $\omega=2J$ and switches branches at $\omega=0$. 
    \item It isn't clear that our function goes to zero fast enough.
    \item $\log(z)$ isn't analytic, it has a singularity at $z=0$, and a branch cut coming out of it.
    \item Some of the entries in the sum have been deleted because they are zero modes.
    \item We have a discrete sum which isn't technically an integral.

\end{enumerate}
Firstly, the discontinuity turns out not to matter. This is because for the wormhole solution the $11$ and $22$ $\Sigma$s choose opposite signs for the square root. This means that we can just switch contours (perfectly fine when performing a sum) and still have a function analytic along the contour, which the branch cut of $\Sigma$ now either entirely above or entirely below it.
\begin{figure}
    \centering
    \includegraphics[scale=0.3]{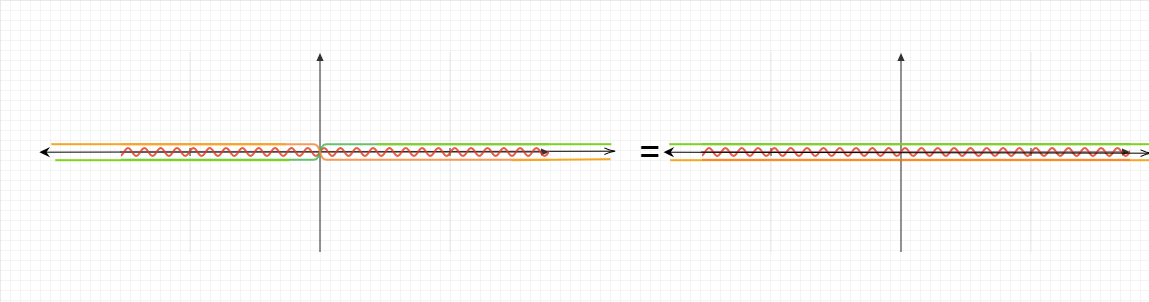}
    \caption{The sum of the contours on the left, where the two contours each cross over the branch cut, is equal to the sum of contours on the right.}
    \label{fig:ComplexContours}
\end{figure}

The second issue is resolved by a cancellation. Let's look at a specific $\omega_1$ and let $\omega_2$ vary. $\log \left( 1-\frac{1}{J^2} \Sigma_{\alpha\alpha}(\omega_1)\Sigma_{\beta\beta}(\omega_2)\right)$ goes to zero as $\Sigma_{\alpha\alpha}(\omega_1)/\omega_2$ as $\omega_2$ goes to infinity. This isn't quite fast enough that we can loop around without changing the integral. But since $\Sigma_{11}(\omega_1)=-\Sigma_{22}(-\omega_1)$, we can close both loops in the complex plane at once.

The third issue is solved by the fourth. We have manually deleted every term that touches the singularity. So evaluating the sum is just a matter of evaluating the difference between the sum and the integral near these deletions.

The fifth issue isn't an issue at all for $|\omega|>2J, JT\gg 1$. In this case the function varies slowly enough that you can replace the sum with an integral with impunity. It is only near the branch cut that we need to be careful. Fortunately, even near the branch cut we can evaluate the difference between the sum and integral exactly for $JT\gg 1$. 



\subsection{F. Transition to linear ramp}
As discussed in the main text, the vacuum manifold shrinks a lot for nonvanishing interactions. 
There is still the isolated saddle point where the two contours are uncorrelated. This point no longer has zero action, and instead takes negative action for small $T$, dominating the path integral. It is also invariant under the time-translation symmetry. There are also the wormhole solutions. They are also found along the original manifold of SU(2) solutions. They all have zero action, and spontaneously break the relative time-translation symmetry.

There will be a transition to chaos which occurs at a time long enough for $G$ to exponentially decay. We can estimate the transition for $J_q\ll J$, where . What is the decay rate? Well, at some $\omega_{\rm crit}$ in the complex plane, the solution to the Schwinger-Dyson equation becomes singular. The decay rate is given by $\textrm {Im} \omega_{crit}$

To first order in $J_q^2$, we can treat the new SD equation as 
\begin{equation}(\omega+J_q^2 \Sigma_q(q))\Sigma=\Sigma^2+J^2
\end{equation}
Here $\Sigma_q$ is the contribution to $\Sigma$ from the $J^2_q$ term. It is a matrix. It is a real odd function of $\omega$ along the diagonals, and a real even function of $\omega$ on the off diagonals, with $\Sigma_{q12}=-\Sigma_{q21}$. This implies a decay rate of $J_q^2/J$.

\subsection{G. A direct evaluation from time translation invariant configurations}
We are going to use a little different notation that is more suitable for the explicit evaluation in this section. Namely, we use $\psi_R \rightarrow -i \psi_R$, and $\hat \Sigma \rightarrow -i \hat \Sigma$. With these modification, the path integral representation of the SFF~\cite{Stanford2018} is
\bea\label{eq:Q-actionS}
    g(T) &=& \int DG D\Sigma \exp N \Big[ \frac12 \Tr \log( \partial_t - \hat\Sigma) - \frac{1}2 \int dt_1 dt_2 (\Sigma_{\alpha \beta} G_{\alpha\beta} - \frac{J^2 }{q}  t_{\alpha\beta} G_{\alpha\beta}^q ) \Big] , 
\label{eq:Action}
\eea
where $t_{\alpha\beta}= \left( \ba{cccc} -1 & i^q \\ i^q & -1 \ea\right)$, and a hat above a variable signals a matrix representation, $(\hat \Sigma)_{\alpha\beta} \equiv \Sigma_{\alpha\beta}$. 
In the case of the SYK$_2$ model, i.e., $q=2$, we can further integrate out $G$ to get
\bea\label{eq:2-actionS}
    g(T) = \int D \Sigma \exp N \Big[ && \frac12 \Tr \log (\partial_t - \hat\Sigma) + \frac1{4J^2} \int \Sigma_{\alpha\beta}^2(t_1, t_2) \Big].
\eea

To attack this horrible path integral, we need to determine stationary points of the action. For general configurations depending on both $t_1-t_2$ and $t_1+t_2$, this is a daunting task. We proceed by assuming that time-translation invariant configurations dominate, namely, $\hat \Sigma(t_1, t_2) =\hat \Sigma(t_1- t_2)$. Taking into account the anti-periodicity around the thermal circle, all fields can then be expanded in the frequency basis. The relevant Matsubara frequencies are $\omega_n = \frac{n \pi}{T}$ with $n$ an odd integer. So long as the dominate saddles are time-translation invariant, configurations that depend non-trivially on $t_1+t_2$ merely contribute fluctuations that do not modify the exponential ramp. 

By the assumption of time-translation symmetry, the variables with different Matsubara frequencies decouple. Thus, in terms of the Fourier component, i.e., $\hat \Sigma_n = \int_0^T dt \hat \Sigma(t) e^{- i \omega_n t}$, the saddle point action becomes
\bea
	g(T) &=& \prod_{n>0, \rm odd} g_n(T), \label{eq:product} \quad 
	g_n(T) 
	= \Big(\frac{N}{\pi} \Big)^2 \int d \sigma_n e^{I[\hat \sigma_n]},\label{eq:gn} \\
	I[\hat \sigma_n] &=& N \Big[ \frac12 \Tr \log (1 - i \frac{\hat\sigma_n}{x_n}) (1 - i \frac{\hat\sigma_n^T}{x_n})  - \frac1{2} \Tr \hat\sigma^2_n \Big], \label{eq:In}
\eea
where we defined dimensionless variables $\hat \sigma_n = \frac1J \hat \Sigma_n$, and $x_n = \frac{\omega_n}{J}$. 
The product over frequency in Eq.~(\ref{eq:product}) is restricted to positive odd integers as we use the symmetry $ \hat \sigma_n = - \hat \sigma_{-n}^T$. 
In Eq.~(\ref{eq:gn}), we implement the normalization by dividing the unnormalized SFF by a free $J=0$ path integral, which evaluates to be $\int d \sigma_n e^{- \frac{N}{2} \Tr \hat\sigma^2_n} = (\frac{\pi}{N})^{2}$. 

A salient feature from the normalization is that one should consider $\hat \sigma_n$ as a hermitian-matrix variable. 
There are two dimensionful variables in this path integral, $J$ and $T$, while the spectral form factor is dimensionless.
This implies they must appear as a product $JT$, so the above normalization gives the correct answer $g(0) = 1$. 
Since we have represented the SFF in Eq.~(\ref{eq:product}) as an infinite product over all positive Matsubara frequencies, the next step is to evaluate $g_n$ at a fixed frequency.

The steepest descent method is accurate because of the large-$N$ structure in Eq.~(\ref{eq:In}). At the first step, we evaluate the saddle point solutoins. The equation of motion from Eq.~(\ref{eq:In}) is
\bea
	\Big(1 - \frac{i\hat\sigma}x \Big)^{-1} = i x \hat\sigma,
\eea
where we abbreviate $x = x_n$ for simplicity. The diagonal solutions are
\bea
	\sigma_\pm &=& \frac12 \big( -i x \pm i \sgn(x) \sqrt{x^2-4} \big),\\
	\hat \sigma^{s s'} &=&  \left( \ba{cccc} \sigma_s & 0 \\ 0 & \sigma_{s'} \ea \right), \quad s, s' = \pm.
\eea
The $\sigma_+$ solution dominates in equilibrium physics as it can be analytically continued to $J=0$ where the self-energy vanishes. But we will see in the following, the $\sigma_-$ solution plays a dispensable role in the SFF.

A crucial feature of the equation of motion is that it is invariant under the SU(2) rotation of a solution,
\bea 
    \hat \sigma \rightarrow U^\dag \hat \sigma U.
\eea
The equation of motion is actually invariant under adjoint transformation by arbitray invertible matrices, but the hermitian property restricts the invertible matrix down to SU(2) rotation. 
Among these four saddle points, $\hat \sigma^{++}$ and $\hat \sigma^{--}$ are invariant under the SU(2) rotation, so they represent isolated points in the configuration space. 
On the other hand, the saddle point $\hat \sigma^{+-}$ spans a degenerate manifold under SU(2) rotation, i.e., there are many degenerate zero modes for each frequency. 
Note that the saddle point $\hat \sigma^{-+}$  can be obtained from $\hat \sigma^{+-}$ by SU(2) rotation, so they live in the same connected manifold.

The on-shell actions $I^{ss'}_x \equiv I[\hat \sigma^{ss'}_{x_n}]$ for four saddle points at a fixed frequency are given by,
\bea
	\frac{I^{ss}_x}N &=& \log \Big[ \frac12(1+ \sqrt{1- s\frac4{x^2}})- \frac1{x^2} \Big] + \frac14 (x- s \sqrt{x^2-4})^2, \nn \\
	\frac{I^{+-}_x}N &=& \frac{I^{-+}_x}N = \log \frac1{x^2} + \frac12(x^2-2).
\eea
When $x>2$, all on-shell actions are real, we need to determine which saddle points will contribute to the SFF via steepest descent method. However, when $x<2$, only $I^{+-}_x$ is real, and the other on-shell actions have imaginary components. This means the contributions from these imaginary component are suppressed compared to $I^{+-}_x$.

To furnish the steepest descent method, we need to analyse fluctuations around saddle points.
Making a small variation to the saddle point solution $\hat \sigma^{ss'} \rightarrow \hat \sigma^{ss'} + \delta \hat\sigma$, the quadratic fluctuation around the saddle point is 
\bea
	\frac{\delta I^{ss'}}N &=& - \frac12 \Tr( \hat \sigma^{ss'} \delta \hat \sigma \hat \sigma^{ss'} \delta \hat \sigma) - \frac12 \Tr \delta \hat \sigma^2.
\eea
In the following, we will divide our discussion about saddle-point fluctuations into two regimes, the large frequencies $x>2$ and the small frequencies $x<2$.

\begin{figure}
\subfigure[]{\label{saddle0}
	\includegraphics[width=5cm]{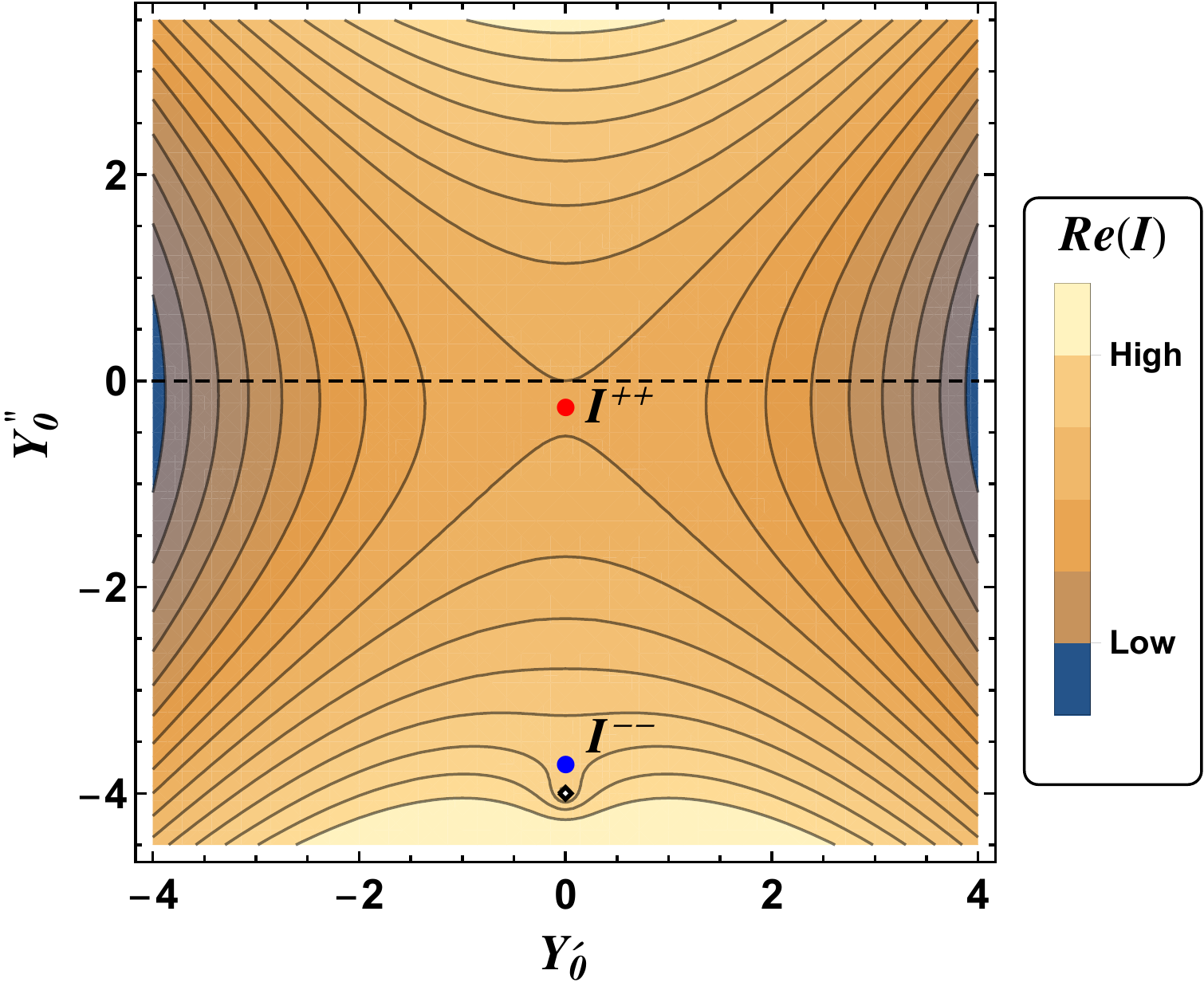}} \quad
\subfigure[]{\label{saddle1}	
	\includegraphics[width=4.05cm]{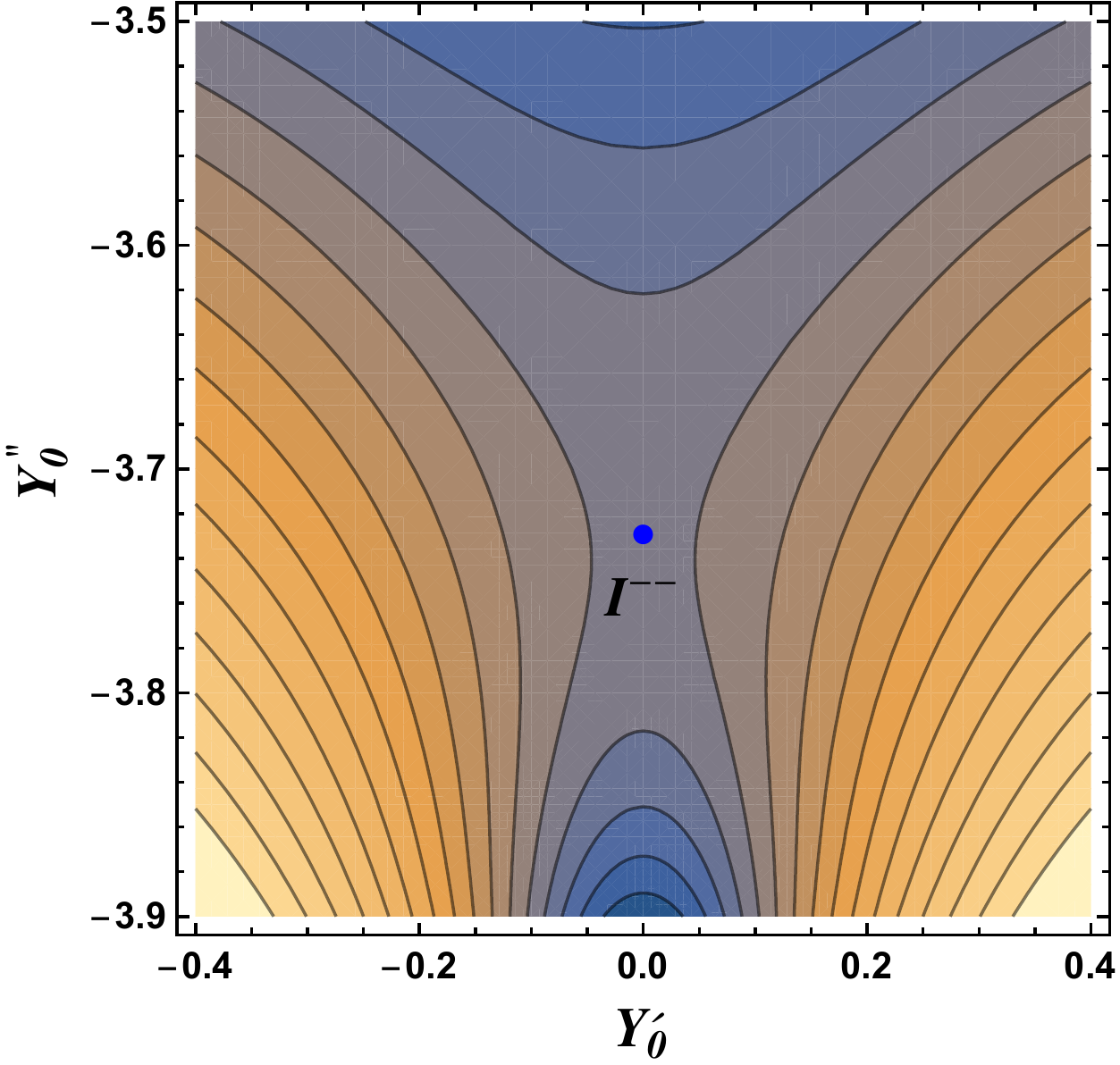}} \quad
\subfigure[]{\label{saddle2}	
	\includegraphics[width=3.95cm]{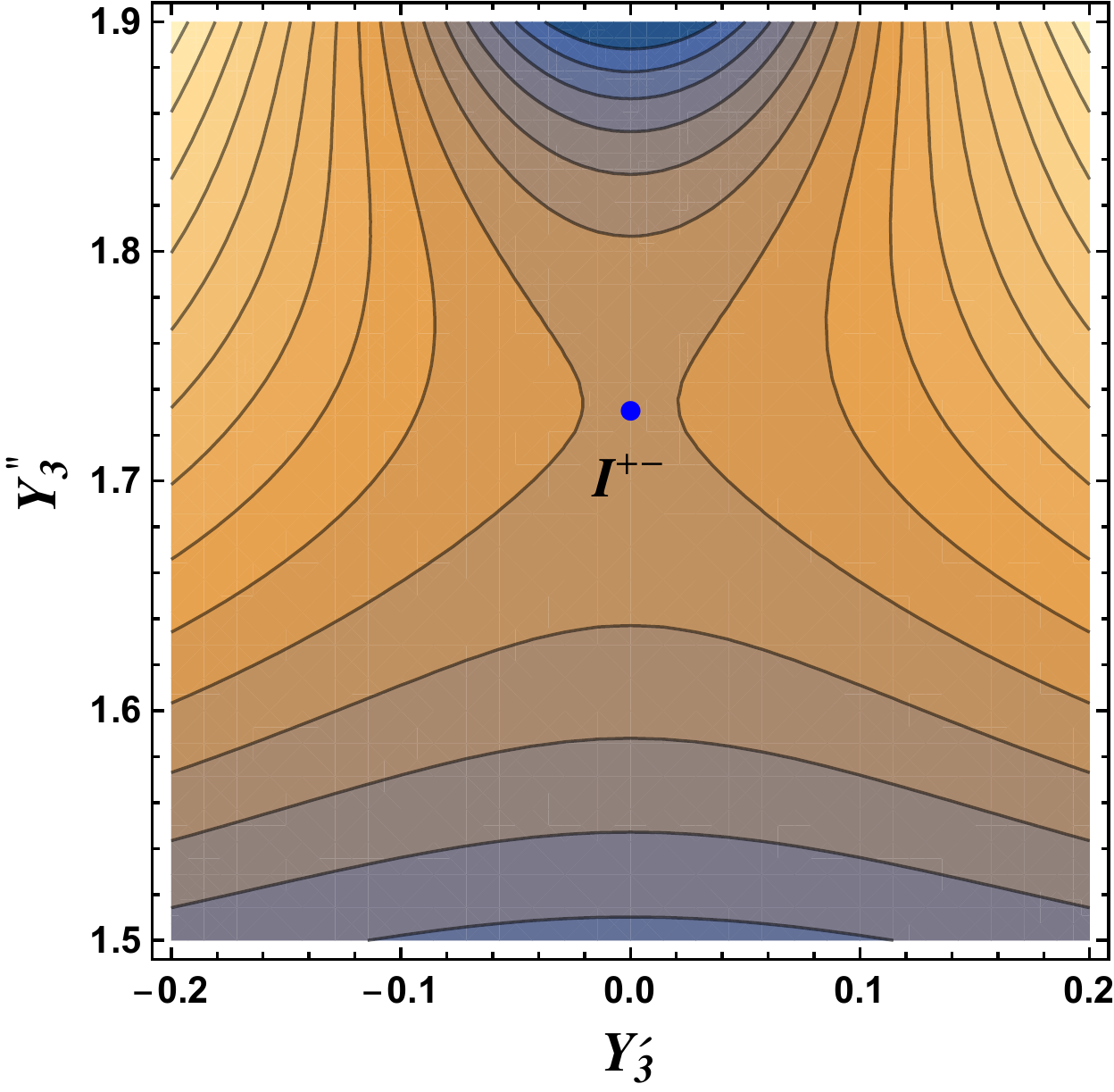}}
	\caption{\label{fig:saddle} The real part of the action near saddle points when $x>2$. The horizontal/vertical axis describes the real/imaginary part of $Y_\mu$, here we plot $\mu=0, 3$ for illustrations.  Notice our definition of action is $e^{I(Y_\mu)}$. The red/blue point indicates the saddle points that can/cannot contribute to the result in the steepest descent method. The original integration original contour is along the horizontal line as indicated by the dashed line in (a). (b) zooms in the saddle point $\sigma^{--}$ in (a). (c) corresponds to the saddle point $\sigma^{+-}$. This also justifies the saddle point solution used in the main text is the right one.}
\end{figure}

At large frequencies, i.e. $x>2$, the quadratic fluctuation around the saddle point $\hat \sigma^{ss}$, $s=\pm$, is
\bea
	\frac{\delta I^{ss}_x}N = -\frac{m^2_{ss}}2  \sum_\mu Y_\mu^2, \quad s=\pm,
\eea
where we have used the Pauli matrix $\sigma^\mu$ to decompose the hermitian matrix $\delta \hat \sigma = \sum_{\mu=0}^4 Y_\mu \sigma^\mu$, $Y_\mu$ are real variables. The effective mass for these two saddle-point fluctuations is given by
\bea
    m^2_{ss} = 4 - x(x - s \sqrt{x^2-4}), \quad s=\pm.
\eea
It is easy to see that $m^2_{++}>0$ and $m^2_{--}<0$ when $x>2$, which indicates the saddle point $\hat \sigma^{--}$ does not contribute in the steepest descent method. 

The the quadratic fluctuation around the saddle point $\hat \sigma^{+-}$, is
\bea
    \frac{\delta I^{+-}_x}N = -\frac{m^2_{++}}2 X_0^2 - \frac{m^2_{--}}2 X_3^2, 
\eea
where $X_{0,3}$ are linear combinations of $Y_{0,3}$. The quadratic fluctuation is independent of $Y_{1,2}$ because there exists degenerate modes. Apparently, because $m^2_{--}<0$, the saddle point $\hat \sigma^{+-}$ cannot contribute to the integral either.

To further show that only the saddle point $\hat \sigma^{++}$ contributes, we plot the real part of the action in Fig.~\ref{fig:saddle}. The negative mass directions of both $\delta I^{--}$ and $\delta I^{+-}$ are perpendicular to the original contour, which means it is impossible to deform the original contour to take account the these two saddle point. Finally, it is a simple task to evaluate the Gaussian integral about the saddle point $\hat \sigma^{++}$ for $x>2$,
\bea
    g_x(T) &=& e^{I^{++}_x}\Big(\frac{N}{\pi} \Big)^2 \int \prod_\mu dY_\mu e^{-m^2_{++} \sum_\mu Y_\mu^2} =\frac{4 e^{I^{++}_x} }{(4-x (x-\sqrt{x^2-4}))^2}.
\eea

At small frequencies, i.e., $0<x<2$, the quadratic fluctuation around the saddle point $\hat \sigma^{+-}$ is
\bea
    \frac{\delta I^{+-}}N \!=\! -\frac{1}2 [ (4-x^2) (Y_0^2 + Y_3^2) - 2i x \sqrt{4-x^2} Y_0 Y_3].
\eea
As a result of the existence of degenerate modes, the quadratic fluctuation has two flat directions $Y_1$ and $Y_2$. Although the mass matrix of $Y_{0,3}$ are complex, fortunately, the real parts of the eigenvalues for this mass matrix are all positive. 
This renders a meaningful steepest descent calculation from the saddle point $\hat \sigma^{+-}$.
\bea
	 g_x(T) &=& e^{I^{+-}_x}\Big(\frac{N}{\pi} \Big)^2 \int \prod_\mu dY_\mu  e^{\delta I^{+-}} = \frac{e^{I^{+-}_x}}{\sqrt{4-x^2}} \frac{N}{\pi} \int dY_1 dY_2 = N \frac{4e^{I^{+-}_x}}{\sqrt{4-x^2}} 
\eea
The last step is from the huge numbers of degenerate zero modes: since the degenerate manifold is a sphere $S^2$, $\int dY_1 dY_2 = 4\pi $. 

The other two saddle points will not contribute to the SFF according to the following two reasons: 1) the on-shell action is oscillating and its contribution mostly cancels, besides that 2) these contributions are suppressed by $1/N$ compared to the saddle point $\hat \sigma^{+-}$ in which the huge degenerate zero modes leads to the enhancement of order $N$.

Since we have evaluated $g_x(T)$ at each frequencies by using the steepest descent method, including the nontrivial contribution from degenerate solution manifold at small frequencies, the SFF is then given by the following product,
\bea
	g(T) 
	&=& \prod_{0<x<2} N \frac{4e^{I^{+-}_x}}{\sqrt{4-x^2} }\prod_{x>2} \frac{4 e^{I^{++}_x} }{(4+x (\sqrt{x^2-4}-x))^2},
\eea
where $x= x_n = \frac{n\pi}{J T}$, $n$ is an odd integer. When $JT \gg 1$, one can approximate the discrete variable $x_n$ by a continuous one, and the SFF is given by
\bea
    g(T) &=& \exp \frac{J T}{2\pi} \Big[ \int_0^2 dx \Big( \log \frac{4 N }{\sqrt{4-x^2}} + I^{+-}_x \Big)+ \int_2^\infty dx \Big( \log \frac{4 }{(4+x (\sqrt{x^2-4}-x))^2} + I^{++}_x \Big) \Big]= \Big( \frac{N}{16 e} \Big)^{\frac{J T}{\pi}},
\eea
which is an exponential ramp. There is minor difference in the subleading order $O(\log \frac1N)$ compared to the main text. This is because we neglect time translation broken configurations in this calculation.

On the other hand, when time is small $J T \ll 1$, the discreteness of $x_n$ will be important, especially, if $J T < \frac\pi2$ there is no contributions from $x<2$. So one would not see the exponential ramp at small times. To the leading order in $J T$, the SFF is given by
\bea
    g(T) = \exp\Big(- \sum_{n \in \rm odd }\frac{N}2 \Big(\frac{JT}{n \pi} \Big)^2 \Big) = e^{ -N \big(\frac{JT}{4} \big)^2}.
\eea
This gives rise to the slope in the SFF of SYK$_2$ model, which is consistent with the result in main text.

\end{widetext}

\end{document}